\documentclass[aps,prl,twocolumn]{revtex4}
\usepackage{graphicx}
\usepackage{amsmath}
\begin{document}
\title{Novel electric field effects on Landau levels in Graphene}
\author{Vinu Lukose}
\email{vinu@imsc.res.in}
\author{R. Shankar}
\email{shankar@imsc.res.in}
\author{G. Baskaran}
\email{baskaran@imsc.res.in}
\affiliation{The Institute of Mathematical Sciences,\\
         C.I.T. Campus, Chennai 600 113, India }

\begin{abstract} 

A new effect in graphene in the presence of crossed uniform electric and
magnetic fields is predicted. Landau levels are shown to be modified in
an unexpected fashion by the electric field, leading to a collapse of
the spectrum, when the value of electric to magnetic field ratio exceeds
a certain critical value. Our theoretical results, strikingly different
from the standard $2d$ electron gas, are explained using a `Lorentz
boost', and as an `instability of a relativistic quantum field vacuum'.
It is a remarkable case of emergent relativistic type phenomena in
non-relativistic graphene. We also discuss few possible experimental
consequence.

\end{abstract}

\maketitle

Graphene has been in the forefront of nano electronics and quantum
condensed matter physics in the last couple of years. The mechanical
rigidity, being able to peel off single graphene layer\cite{prep} and
ability to make electrical contacts has made the system very appealing
from building devices and experimental points of view \cite{expt}
\cite{expt-bilayer}. One of the remarkable discoveries in graphene is the
anomalous $2d$ quantized Hall effect. A variety of rich physics and
anomalous phenomena are tied to a remarkable `relativistic like'
spectrum that electron and holes possess in graphene \cite{theory}
\cite{theory-bi}. This has made graphene important and interesting from
several points of view in physics.

In the present letter, we investigate the effect of a uniform electric
field, applied along the graphene sheet, on its already anomalous Landau
level spectrum. We find that, within the low energy approximation near
the Fermi surface (Dirac points), the problem can be exactly solved.
\emph{We find strikingly new effects of electric field on the Landau
levels which is different from the Landau levels of standard $2d$
electron gas}.

We find that the Landau spectrum gets scaled, for a given $k_y$ quantum
number, by an electric field dependent dimensionless
parameter($\beta=\frac{E}{v_FB}$).  As the value of this parameter is
increased, spacings between the Landau levels decreases. This Landau
level contraction is consequence of electric field induced quantum
mechanical mixing of Landau levels. The entire Landau level structure
collapses at a critical value of this parameter.  Further, the
`relativistic' character of the spectrum (with Fermi velocity replacing
the velocity of light), leads to a novel interpretation of our result in
terms of relativistic boosts and the mixing of electric and magnetic
fields in moving frames of reference.  We confirm our analytical result
by solving the full tight binding model for a graphene sheet in the
presence of magnetic and electric fields numerically.  The collapse seen
in the low energy approximation, is indeed accelerated in the actual
tight binding model. 

The modified wave functions and energy spectrum will have implications
on the nature of quantum Hall break down. We briefly touch upon this
issue at the end, and point out how it could be different from the
standard quantum Hall break down.

Electronic states of graphene are well described by the tight binding
hamiltonian for the $\pi$ electrons of the carbon atoms. In graphene,
the carbon atoms form a triangular lattice with a basis of two
geometrically inequivalent atoms placed $\frac{a}{\sqrt{3}}$ apart,
where $a=2.456\, A^0$ is the lattice constant. The overlap integral
between the nearest carbon atoms is $t \approx 2.71 eV$. We denote the
triangular lattice sites by ${\bf R}_i=i_1\hat{{\bf e}}_1+i_2\hat{{\bf
e}}_2$, where $\hat{{\bf e}}_{1}= \hat{{\bf x}}$ and $\hat{{\bf e}}_{2}
= -\frac{1}{2}\hat{{\bf x}} + \frac{\sqrt 3}{2}\hat{{\bf y}}$ are the
basis vectors.  $c_{{\bf i}r\sigma} ~ (r=1,2 ~ {\rm and} ~
\sigma=\uparrow,\downarrow)$ represent the electron annihilation
operators with sub-lattice index $r$ and spin index $\sigma$ at ${\bf
R}_i$. The hamiltonian is then written as,
\begin{eqnarray} 
H = -t \sum_{{\bf i}\sigma} {c^\dagger}_{{\bf i}2\sigma} \left( 
c_{{\bf i}1\sigma} + c_{{\bf i}+\hat{{\bf e}}_2 1\sigma} 
+ c_{{\bf i}-\hat{{\bf e}}_3 1\sigma} \right) + h.c.
\label{tight-binding} 
\end{eqnarray}
where $\hat{{\bf e}}_3=-(\hat{{\bf e}}_1+\hat{{\bf e}_2})$. The
electronic dispersion for graphene has two points in a Brillouin zone
which separates the positive and negative energy eigenstates.  These so
called Dirac points are $K_{1,2} = \pm \frac{2\pi}{a}
(\frac{1}{\sqrt{3}} \hat{\bf x} + \hat{\bf y})$.  The dispersion
relation in the proximity of the Dirac points is linearly proportional
to $\vert{\bf k} \vert$ . The low energy modes around these points are
described by slowly varying fields $\psi_{r \eta \sigma}({\bf R}_i)$
defined as,
\begin{equation}
c_{ir\sigma} = e^{i{\bf K}_1.{\bf R}_i} \alpha_{rr^\prime}^z
\psi_{r^\prime 1 \sigma}({\bf R}_i) + e^{i{\bf K}_2.{\bf R}_i}
\alpha_{rr^\prime}^x \psi_{r^\prime 2 \sigma}({\bf R}_i)
\end{equation}
Where $\alpha^x,\alpha^y$ are the Pauli matrices. The effective
hamiltonian for the low energy modes is the Dirac hamiltonian.
\begin{equation}
H = v_F \int d^2x~\sum_{\eta\sigma}
\Psi^{\dagger}_{\eta\sigma} ~ {\textrm{\boldmath$\alpha$}}.{\bf p} 
~ \Psi_{\eta\sigma}
\label{continuum ham}
\end{equation}
where $v_F=\frac{\sqrt{3}}{2}\frac{at}{\hbar}$ is the Fermi velocity.
$\Psi_{\eta\sigma}$ are two component field operators where $\eta(=1,2)$
is the valley index, corresponds to two Dirac points and
$\sigma(=\uparrow,\downarrow)$ is the spin index. The spectrum can be
obtained by solving the one particle equation to get the linear
dispersion, $\epsilon({\bf k}) = \pm \hbar v_F|{\bf k}|$.  In presence
of an external magnetic field perpendicular to the graphene plane the
one particle hamiltonian, $h=v_F {\textrm{\boldmath$\alpha$}}.{\bf\Pi}$,
where ${\bf\Pi} = {\bf p} + e{\bf A}$, The energy eigenvalues are
\begin{equation}
\epsilon_{n,k_y} = \operatorname{sgn}(n) \sqrt{2\vert n \vert} 
\frac{\hbar v_F}{l_c}
\label{spectrum ll}
\end{equation}
\begin{figure}
\rotatebox{270}{\includegraphics[height=3.4in]{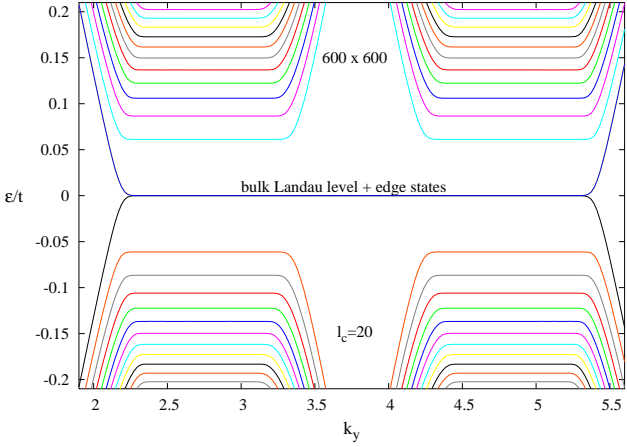}}
\caption{Energy eigenvalues $\epsilon_{n,k_y}$, for electrons in
graphene computed from the tight binding model for a hexagonal lattice
subjected to a  magnetic field $B=27.3$ Tesla (or $l_c=20\,a$ where $a$
is the triangular lattice spacing) for a system size of $600a \times
600a$.  The plot shows $\epsilon_{n,k_y}$ in the units of $t$ as
function of $k_y$, where $k_y$ is the wavevector in the $y$ direction.
Two sets of horizontal lines are Landau levels corresponding to the two
valleys and $n=0$ Landau level and the edge states are degenerate }
\label{plot1}
\end{figure}
$n$ is the Landau level index, $k_y=\frac{2\pi}{L_y} l$ is the quantum
number corresponding to translation symmetry along $y$-axis, both $n$
and $l$ are integers (we choose Landau gauge ${\bf A}({\bf
r})=xB\,\hat{\bf y}$) and $l_c = \sqrt{\frac{\hbar}{eB}} $ is the
magnetic length. Unlike the case of the non-relativistic electron in a 
magnetic field, where the spectrum has a linear dependence on the magnetic field
and the non-negative integer valued Landau level index, the graphene Landau
levels have a square root dependence on both magnetic field and Landau
level index. The degeneracy of each level is given by the number of
magnetic flux quanta passing through the sample. The eigenfunctions are,
\begin{equation}
\psi_{nk_y}(x,y) \propto e^{ik_yy}\left( \begin{array}{c}
\operatorname{sgn}(n)\, \phi_{|n|-1}(\xi)\\
i\,\phi_{|n|}(\xi)
\end{array}\right)
\label{efns1}
\end{equation}
where $\phi_n(\xi)$ are the harmonic oscillator eigen-functions and
$\xi\equiv \frac{1}{l_c}\left(x+l_c^2k_y\right)$.

We now consider the above system in the presence of a constant electric field
in the $x$-direction. The single particle hamiltonian is then given by,
\begin{equation}
h = v_F {\textrm{\boldmath$\alpha$}}.{\bf\Pi}+{\bf{1}}eEx
\label{ham2}
\end{equation}

The Lorentz covariant structure of the hamiltonian, with $v_F$ playing
the role of the speed of light, can be used to solve it exactly
\cite{mac}. It is known from special relativity, if $v_FB>\vert{\bf
E}\vert$, then we can always boost to a frame of reference where the
electric field vanishes and the magnetic field is reduced. We can then
use the solution in Eq.(\ref{spectrum ll}) and boost back to get the
exact spectrum of the hamiltonian in Eq.(\ref{ham2}). Here the
boost transformation amounts to doing a transformation on the space-time
coordinate system. To implement the above procedure, it is convenient to
work with the manifestly covariant time dependent Dirac equation,
\begin{equation}
i \hbar \gamma^{\mu} ( \partial_{\mu} + i \frac{e}{\hbar}A_{\mu} )
    \Psi(x^{\mu}) = 0
    \label{dirac eom}
\end{equation}
\begin{figure}
\rotatebox{270}{\includegraphics[height=3.4in]{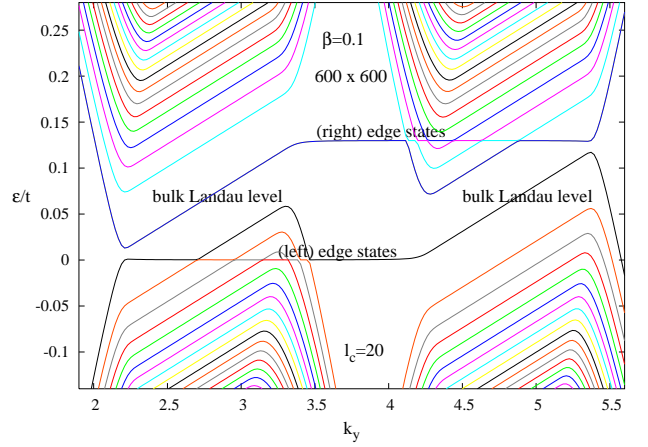}}
\caption{Energy eigenvalues $\epsilon_{n,k_y}$ for electrons computed
for the tight binding model for parameters given in Fig.\ref{plot1} and
an external electric field $E$ applied along $x$-axis, given by the
parameter $\beta=\frac{E}{v_F B} = 0.1$. The electric field gives a
linear $k_y$ dependence to the bulk Landau levels whereas it gives a
constant shift to the edge states. The part of solid line labelled `bulk
Landau level' are $n=0$ Landau levels and parallel lines above and below
them are Landau levels corresponding to positive and negative $n$
respectively. Set of points parallel to $k_y$ labelled `edge states' are
surface states localised at the zig-zag boundary.  }
\label{plot2}
\end{figure}
Where $ x^0 = v_F t$, $x^1 = x$, $x^2 = y $, $ \gamma^0 = \alpha^z$,
$\gamma^1 = i\alpha^y$, $\gamma^2 = -i\alpha^x$,  $\partial_\mu =
\frac{\partial}{\partial x^\mu}$. $A^0 = \phi$, the scalar potential,
$A^1 = A_x$, $A^2 = A_y$ and $\Psi(x^{\mu})$ is a two component spinor.
We  now apply a Lorentz boost in the $y$-direction (perpendicular to the
electric field),
\begin{eqnarray}
 \left(
 \begin{array}{c}
 \tilde{x}^0 \\
 \tilde{x}^2
 \end{array}
 \right)
  =
\left(
 \begin{array}{cc}
 \cosh\theta & \sinh\theta\\
 \sinh\theta & \cosh\theta
 \end{array}
\right)
\left(
 \begin{array}{c}
 {x^0}\\
 {x^2}
 \end{array}
\right) 
\label{boost}
\end{eqnarray}
and $\tilde{x}^1=x^1$. The wave function transforms,
$\tilde{\Psi}(\tilde{x}^\mu) = e^{\frac{\theta}{2}\alpha_y}\Psi(x^\mu)$.
Applying the above transformations and choosing $ \tanh\theta =
\frac{E}{v_F B} = \beta$, we can rewrite the Dirac equation in Eq.(\ref{dirac eom}),
\begin{eqnarray}
\left(\gamma^0 \tilde{\partial}_0 + \gamma^1 \tilde{\partial}_1 + \gamma^2 (
            \tilde{\partial}_2 + \frac{i}{l^2_c} \sqrt{1-\beta^2}\:
            \tilde{x}^1)\right)\tilde{\Psi}(\tilde{x}^\mu)
    = 0 \label{eom2}
\end{eqnarray}

In the boosted coordinates, where $\vert \beta \vert < 1$ ,it is a
problem of a Dirac electron in a (reduced) magnetic field, $\tilde{B} =
B\sqrt{1-\beta^2}$. The time component of the $3$-momentum in the
boosted frame, $\tilde{\epsilon}_{n,\tilde{k}_y} = \operatorname{sgn}(n)
\sqrt{2 \vert n \vert}  \frac{\hbar v_F}{l_c} (1-\beta^2)^{\frac{1}{4}}
$ is not the physical energy eigenvalue of our problem. We have to apply
the inverse boost transformation to obtain the spectrum and
eigenfunctions of our problem,
\begin{equation}
\epsilon_{n,k_y} = \operatorname{sgn}(n) \sqrt{2\vert n \vert} 
\frac{\hbar v_F}{l_c}{(1-\beta^2)}^{\frac{3}{4}} - \hbar v_F \beta k_y 
\label{scaled ll}
\end{equation}
\begin{eqnarray}
\begin{array}{c}
\Psi_{n,k_y}(x,y)
\end{array}
\propto e^{ik_yy}e^{-\frac{\theta}{2} \alpha_y}
\left(
\begin{array}{c}
\operatorname{sgn}(n)\, \phi_{|n|-1}(\xi') \\
i\, \phi_{|n|}(\xi')
\end{array}
\right)
\label{efns2}
\end{eqnarray}
\begin{equation}
\xi' \equiv \frac{(1-\beta^2)^\frac{1}{4}}{l_c}\left( x+l_c^2 k_y
+ \operatorname{sgn}(n)\frac{\sqrt{2\vert n\vert }\: l_c \beta}
{(1-\beta^2)^\frac{1}{4}}\right) 
\label{efncenter}
\end{equation}

The energy eigenvalues of the standard $2d$ electron gas in crossed
magnetic and electric fields are given by $\epsilon_{n,k_y} =
(n+\frac{1}{2})\hbar \omega_c - \hbar k_y \frac{E}{B} -
\frac{m}{2}(\frac{E}{B})^2$. The main difference between the two besides
the $\sqrt{n}$ and $\sqrt{B}$ dependence, is that \emph{the low lying
graphene Landau level spacing scales as $(1-\beta^2)^{\frac{3}{4}}$},
whereas the spacing is independent of the electric field in the
non-relativistic case.  Comparing the eigenfunctions with and without
the electric field (\ref{efns2}, \ref{efns1}), we see that the effect of
the electric field is to (un)squeeze the oscillator states as well as to
mix the particle and hole wave-functions. Squeezing corresponds to the
change in $l_c$ and the eigenfunctions in Eq.(\ref{efns2}) can be
expanded as superposition of states in Eq.(\ref{efns1}). \emph{Thus,
unlike in the usual semiconductor samples, in graphene the electric
field causes Landau level mixing}. Also notice in (\ref{efncenter}) that
\emph{the location of the gaussian also shifts as a function of the
Landau level index n}, unlike the standard $2d$ electron gas. 

As $\beta$ approaches unity, from Eq.(\ref{efncenter}) we infer that,
to keep the gaussian shifts within the linear extent of the system
requires larger values of $k_y$, which takes us beyond the long
wavelength approximation. Moreover the Eq.(\ref{scaled ll}) hands a
collapse of the Landau level spectrum at $\beta=1$. One may wonder if
the collapse we have found is an artifact of the low energy
approximation?  Interestingly, we find that in our full tight binding
calculation the collapse persists, and infact it occurs at a value of
$\beta$ even smaller than unity. 

We have performed extensive numerical computations on the tight binding
model for graphene with magnetic and electric fields, using lattice
sizes ranging from $60\times60$ to $600\times600$. The magnetic field
enters through the Peierls substitution, $t \to t\:e^{i \frac{2\pi
e}{\hbar} \int{\bf A}.d{\bf l}}$ in Eq~(\ref{tight-binding}).  ${\bf
A}({\bf r})$ is chosen in such a way that the contribution to the phase
term comes from hopping along one of the three bonds for each carbon
atom. This enables us to maintain translation symmetry along the
$\hat{\bf e}_2$ axis of the triangular lattice. The problem then reduces
to the $1$D Harper equation.
\begin{eqnarray}
\epsilon \phi_{1,n_1} = 2t \cos(\frac{k_2 a + n_1\varphi}{2})
        \phi_{2,n_1} + t \phi_{2,n_1+1} \nonumber \\
\epsilon \phi_{2,n_1} = 2t \cos(\frac{k_2 a + n_1\varphi}{2})
        \phi_{1,n_1} + t \phi_{1,n_1-1}
\end{eqnarray}
Here $\varphi$ is the magnetic flux passing through each plaquette,
$k_2$ is the wave vector and $n_1$ is the $\hat{\bf e}_1$ component of
triangular lattice coordinate.
\begin{figure}
\begin{center}
\begin{tabular}{ll}
\includegraphics[width=1.6in,height=2.4in]{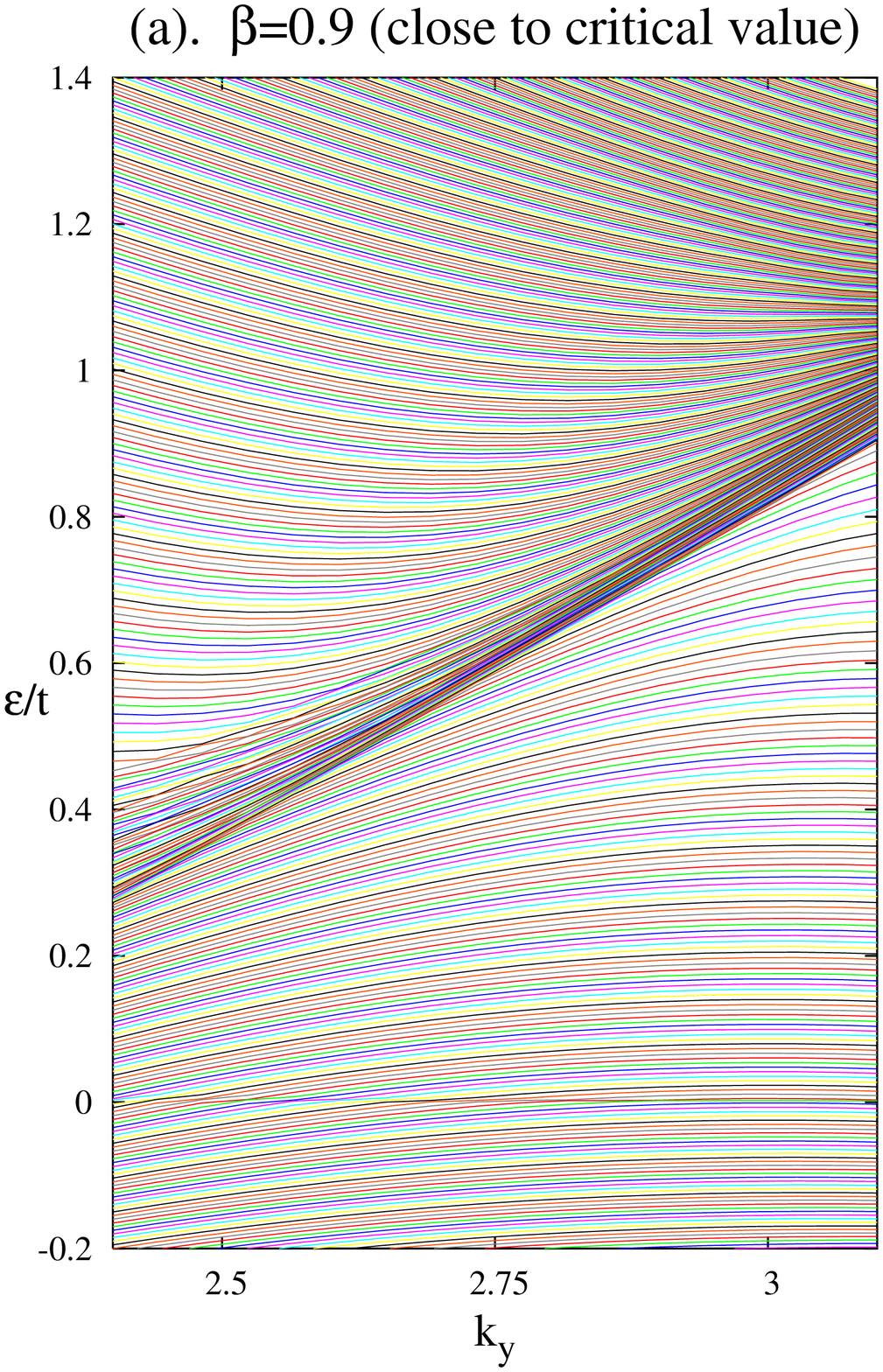} &
\includegraphics[width=1.6in,height=2.4in]{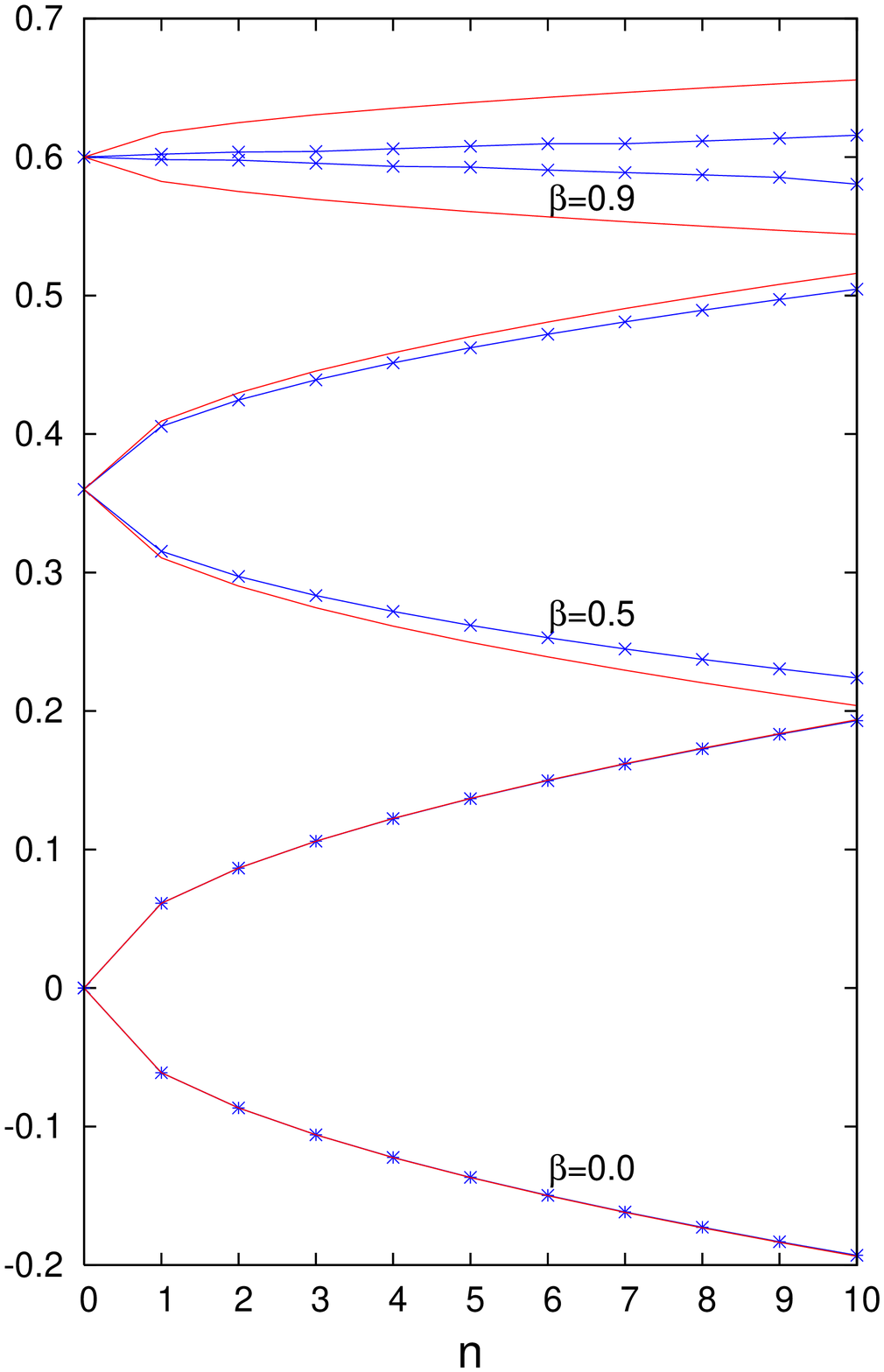}
\end{tabular}
\caption{(a) shows the energy eigenvalues around the Dirac point plotted
as function of $k_y$ for $\beta=0.9$. The collapse can be clearly seen
(b) shows the modulus of the eigenvalues $|\epsilon_{n,k_y}|$ for a
value of $k_y=2.785$ computed from the tight binding model for system
size $600a \times 600a$, magnetic field $B=27.3\: Tesla$ or $l_c=20\,a$,
electric field given by parameter $\beta = 0.0,0.5,0.9$, plotted as a
function of $n$.}
\label{plot3}
\end{center}
\end{figure}

We choose the value of the magnetic field such that $L \gg l_c \gg a$,
where $L$ is the linear extent of the system. The condition $l_c \gg a $
ensures that we stay away from the Hofstadter butterfly kind of
commensurability effects on the spectrum and $L \gg l_c$ ensures that a
large number of cyclotrons orbits fit in the sample. For our numerics,
we expressed all energies in units of $t$ and all lengths in units of
$a$.

Fig.(\ref{plot1}) shows the results of our numerical investigation for
zero and Fig.(\ref{plot2}) for a finite ($\beta = 0.1$) electric fields.
Fig.(\ref{plot1}) shows the spectrum at low energies and the eigenvalues
that are constant w.r.t $k_y$ are the Landau levels. They have
$\sqrt{n}$ behaviour and are in excellent agreement with the analytical result
and the eigenstates that vary with $k_y$ are the chiral edge states
responsible for the quantum Hall current. In our numerics the lattice has
zig-zag edges at the two ends along $\hat{\bf e}_1$.  It is well known
that the zig-zag edges result in zero-eigenvalue states even in the absence
of external magnetic field \cite{edge}. These zero-eigenvalues surface
states which are localised at the boundary and $n=0$ Landau level forms
degenerate set of states as shown in Fig.(\ref{plot1}).  However,
Fig.(\ref{plot2}) shows that this degeneracy gets lifted in presence of
an electric field. For small electric fields the wavefuntion of these edge
states continues to be localised near edge of the sample.  Any difference
in eigenvalues of these surface states is because of the potential seen by
them due to the externally applied electric field at the two edges of the
sample. A characteristic feature of these edge states is that they don't
vary with the wavevector, whereas the Landau levels develop a linear
$k_y$ dependence with electric field. 

Fig.(\ref{plot3}b), shows  ${\sqrt n}$ scaling of Landau levels for a
given $k_y$ value. For zero electric field we see an excellent match
between analytics and numerics. And for the case of finite electric
fields we see a systematic deviation from exact results as we suspected
from our exact result. As $\beta \rightarrow 1$, the tight binding
results shows a faster collapse. Fig.(\ref{plot3}a) shows the collapse
has already occurred at $\beta = 0.9$, near one the Dirac points. 

We show below that one of the consequences of the Landau level
contraction (\ref{scaled ll}) and the $n$ dependent guassian shift
(\ref{efncenter}) is the possibility of a `dielectric breakdown', which
is different from the conventional ones. The single particle spectrum
and states we have obtained thus far (for a given $E$ and $B$) can be
used to construct stable many-body quantum Hall ground states. However,
the external electric field not only modifies the single particle wave
function and spectrum, but can also destabilise the ground state through
spontaneous creation of particle-hole pairs; i.e., by a dielectric
breakdown. We present a simple formula for dielectric breakdown, without
giving full details. It has an unusual dependence on the length scale
over which the potential fluctuates and on the Landau level index $n$.
This peculiar feature is absent in standard quantum $2d$ electron
systems \cite{break}.  Specifically we find that for slowly varying
electric field fluctuations over a length scale $\ell_E$ and for large
Landau level index, the critical voltage for breakdown is given by:
\begin{equation}
V_{c}\approx \frac{\Delta_n(0,B)}{e}\left( 1 \pm \kappa\, n \left(
\frac{l_c}{\ell_E} \right)^2 \right)
\label{vcritsol}
\end{equation}
where $\kappa$ is a constant of the order of unity, which depends on the
strength of the electric field fluctuations and $\Delta_n$ is gap
between levels $n$ and $n+1$. This means that if we have
an electric field, non-uniform over a nanoscopic scale ($\ell_E$ $\sim
l_c$), it will cause local breakdown even before the critical field is
reached. Such situations can be created through in plane or out of plane
charged impurities or STM tips, in addition to external electric fields.
It is interesting that such an anomalous local breakdown is Landau level
index $n$ dependent, we expect that the quantum Hall breakdown should
be qualitatively different for $n=0$ and $n\neq0$ within graphene.

In the light of new spectroscopic experiments \cite{heer}, we claim that
the contraction in Landau level spacing and the collapse can be observed at
fields attainable in laboratories. The gap between $n=0$ and $n=1$ for
$B\sim 1\,$Tesla is $\sim 35$ meV, for $E\sim3\times10^5\,Vm^{-1}$,
$10\%$ reduction in the gap is expected. And the collapse of the Landau
levels  should also be observeable by applying $E \sim 10^6\,Vm^{-1}$.
In the context of quantum Hall breakdown, the dependence of critical
voltage on  $\ell_E$ as given in Eq.(\ref{vcritsol}) suggests that the
breakdown phenomena should be different from what we observe in
standard $2d$ quantum Hall system.  Moreover graphene's Landau level
index dependence on $V_c$, we expect the breakdown phenomena is going
to be different for $n\neq0$ from that of $n=0$.

It will be interesting to study graphene from the point of view of the
present paper. As quantum Hall phenomena are beginning to be seen in
pyrolytic graphite\cite{kopelevich} and possibly in carbon
eggshells\cite{timir}, it will be very interesting to study electric
field effects in these systems as well, to confirm our predictions.

In summary, we have made a theoretical prediction of a remarkable
phenomena in graphene: Landau level contraction and an eventual
collapse, induced by crossed electric fields. The local dielectric
breakdown has a peculiar length and Landau level index dependence. These
phenomena, not known in the standard $2d$ electron gas, is a  consequence of
the relativistic type spectrum of low energy electrons and holes in
graphene.


\begin{references}
\bibitem{prep} K.S. Novoselov et al., Science, {\bf 306}, 666(2004); C.
Berger et al., J. Phys. Chem., {\bf 108}, 19912 (2004)
\bibitem{expt} K.S. Novoselov et al., Nature {\bf 438},197(2005); Y.
Zhang et al., Nature {\bf 438},201(2005);  Y. Zhang et al., Phys. Rev.
Lett. {\bf 96}, 136806 (2006)
\bibitem{expt-bilayer} K.S. Novoselov et al.,Nature Physics {\bf 2},
177-180(2006)
\bibitem{theory} V.P. Gusynin, S.G. Sharapov, Phys.Rev.Lett. {\bf 95}
146801(2005); C.L. Kane, E.J. Mele, Phys. Rev. Lett. {\bf 95}
146802(2005); ibid {\bf 95}, 226801(2005); A. H. Castro Neto, F. Guinea,
N. M. R. Peres, Phys.Rev.B {\bf 73}, 205408 (2006); M. I. Katsnelson,
cond-mat/0512337; J.  Tworzydlo et al.,cond-mat/0603315;
D.V.Khveshchenko, cond-mat/0602398; N. A.  Sinitsyn et al.
cond-mat/0602598; D. N. Sheng, L. Sheng, Z. Y. Weng cond-mat/0602190; N.
M. R. Peres, F. Guinea, A. H. Castro Neto, Phys.Rev. B {\bf 73},
125411(2006); ibid {\bf 72}, 174406(2005);ibid. {\bf 73},125411
(2006),cond-mat/0512476; cond-mat/0506709; cond-mat/0603155
\bibitem{theory-bi}E.McCann, V. I. Fal\'ko, Phys.Rev.Lett.{\bf 96},
086805(2006); J. Nilsson et al cond-mat/0512360;
\bibitem{mac} A.H. MacDonald, Phys. Rev. B {\bf 28}, 2235(1983)
\bibitem{edge}Y. Niimi et al., Phys. Rev. B {\bf 73}, 085421(2006); L.
Brey, H.A. Fertig, cond-mat/0602505; cond-mat/0603107; K. Sasaki, S.
Murakami, R.  Saito, cond-mat/0602647; A. Abanin, P. A. Lee, L. S.
Levitov, Phys. Rev. Lett.  {\bf 96}, 176803(2006); D. N. Sheng, L.
Sheng, Z. Y. Weng, cond-mat/0602190; V.  M. Pereira et al, Phys. Rev.
Lett. {\bf 96}, 036801 (2006)
\bibitem{break} G. Ebart et al., J. Phys. C {\bf 16}, 5441(1983); V.
Tsemekhman et al., Phys. Rev. B {\bf 55}, R10201
\bibitem{heer} M.L. Sadowski et al., cond-mat/0605739
\bibitem{kopelevich} H. Kempa, P. Esquinazi, Y. Kopelevich,
cond-mat/0603155
\bibitem{timir}Timir Datta, et al cond-mat/0503166.
\end{references}
\end{document}